\documentclass[prb,twocolumn]{revtex4}
\usepackage{graphicx}
\usepackage{hyperref}
\usepackage{amssymb}
\usepackage{caption}
\usepackage{dcolumn}
\usepackage{float}
\usepackage{booktabs}
\usepackage{amsmath} 
\usepackage[utf8]{inputenc}
\usepackage{lmodern}
\usepackage{color}
\definecolor{red}{rgb}{1.0,0.0,0.0}

\newcommand{\nn}{\nonumber}

\newcommand{\e}{\mathrm{e}}

\newcommand{\bs}{\boldsymbol}
\DeclareMathAlphabet{\bi}{OML}{cmm}{b}{it}

\def\ba{\begin{aligned}}
\def\ea{\end{aligned}}
\def\be{\begin{equation}}
\def\ee{\end{equation}}
\def\bearr{\begin{eqnarray}}
\def\eearr{\end{eqnarray}}

\def\l{\left}
\def\r{\right}

\usepackage{tikz}

\begin{document}
\title{Floquet topological phase transition in $\alpha$-$\mathcal{T}_3$ lattice}
\bigskip
\author{Bashab Dey and Tarun Kanti Ghosh\\
\normalsize
Department of Physics, Indian Institute of Technology-Kanpur,
Kanpur-208 016, India}
\begin{abstract}
We investigate topological characteristics of the photon-dressed band structure
of $\alpha$-$\mathcal{T}_3$ lattice on being driven by off-resonant 
circularly polarized radiation.
We obtain exact analytical expressions of the quasienergy bands 
over the first Brillouin zone. 
The broken time-reversal symmetry caused by the circularly polarized light lifts
the triple point degeneracy completely at both the Dirac points.
The gaps become unequal at $ {\bf K}$ and ${\bf K}^{\prime}$ (except at $\alpha=0$ and 1), which 
reveals the absence of inversion symmetry in the system.
At $\alpha=1/\sqrt{2}$, the gap between flat and valence bands closes at ${\bf K}$, 
while that between conduction and flat bands closes at  ${\bf K}^{\prime}$, thereby restoring 
a semimetalic phase. At the gap closing point ($\alpha=1/\sqrt{2}$) which is independent 
of the radiation amplitude, there is a reappearance of low-energy Dirac cones
around ${\bf K}$ and ${\bf K}^{\prime}$ points.
Under the influence of the circularly polarized radiation, the $\alpha$-$\mathcal{T}_3$ lattice 
is transformed from semimetal to a Haldane-like Chern insulator
characterized by non-zero Chern number.
The system undergoes a topological phase transition from $\mathcal{C} = 1 (-1)$ to 
$\mathcal{C}=2 (-2)$ at $\alpha =1/\sqrt{2}$, where $\mathcal{C}$ is the Chern number 
of the valence (conduction) band.
This sets an example of a multiband system having larger Chern number. 
These results are supported by the appearance of chiral edge states in irradiated 
$\alpha$-$\mathcal{T}_3$ nanoribbon.

\end{abstract}

\maketitle

\section{Introduction}
Non-trivial topological phases in 
electronic and photonic systems have drawn enormous interest 
since the discovery of quantum Hall effect \cite{qhe}. 
Topological insulators (TIs) \cite{TI-rmp1,TI-rmp2,TI-photon} are distinctive states of matter
characterized by an insulating bulk gap and gapless chiral or helical edge/surface
modes that are topologically protected 
\cite{TI-Kane1,TI-Kane2,TI-scz,FDMH,CI1,CI2,CI3,CI4}. 
There are several classes of TIs, each of which is represented by a topological index. 
Chern insulators (CIs),  also known as anomalous quantum Hall insulators (AQHIs) 
belong to a class of TIs characterized by a topological invariant called 
Chern number $\mathcal{C}$ associated with each band. A band with non-zero 
$\mathcal{C}$ gives rise to quantized Hall conductance even in the absence of a 
net magnetic flux. This feature was first predicted in an exotic 2D lattice model 
with broken time-reversal (TRS) symmetry, popularly known as the 
Haldane model\cite{FDMH} and was later verified experimentally \cite{haldane-exp}. 
These materials host chiral edge states (unidirectional propagating 
modes along an edge) that are robust against backscattering. 
The edge states are guaranteed by a non-zero Chern number of the bulk band 
through bulk-edge correspondence\cite{hatsugai}.
On the other hand, $\mathbb{Z}_2$ TIs, also called quantum spin Hall insulators (QSHIs), 
constitute another class of TIs in which the edge states are protected by time-reversal 
symmetry \cite{TI-Kane1,TI-Kane2}. $\mathbb{Z}_2$ phases have been
studied in large number of systems including two-dimensional (2D) quantum 
materials, strong spin-orbit coupled quantum wells and exotic lattice models.
The gapless edge states in $\mathbb{Z}_2$ TIs are  helical i.e. they form pairs 
of counter-propagating modes with opposite spins along an edge, that are time-reversed 
copies of each other. The topology of these edge states is described by another 
topological invariant called the $\mathbb{Z}_2$ index.
The topological phases exist in static \cite{TI-static} as well as 
in time-periodic systems 
\cite{TI-periodic1,TI-periodic2,TI-periodic3,Kitagawa,TI-periodic4,TI-periodic5,TI-periodic6}.
Such a periodic drive can also transform a topologically trivial insulator 
to a CI \cite{usaj,graphene-CI-exp}. The properties of periodically driven systems can be 
analyzed using Floquet theory \cite{Floquet,Michael}.

The Chern number, in principle, can have any integral value.
Most of the theoretically and experimentally studied CIs have Chern 
number $\mathcal{C}=1$. Therefore, it would be interesting to have a system with
Chern number $\mathcal{C} \geq 2$. Recently, large Chern numbers have been predicted \cite{highC-t} 
and experimentally \cite{highC-exp} 
realized in photonic 2D square and hexagonal crystals. 
 
\begin{figure}[htbp]
\includegraphics[trim={0cm 1cm 0cm 5cm},clip,width=8.5cm]{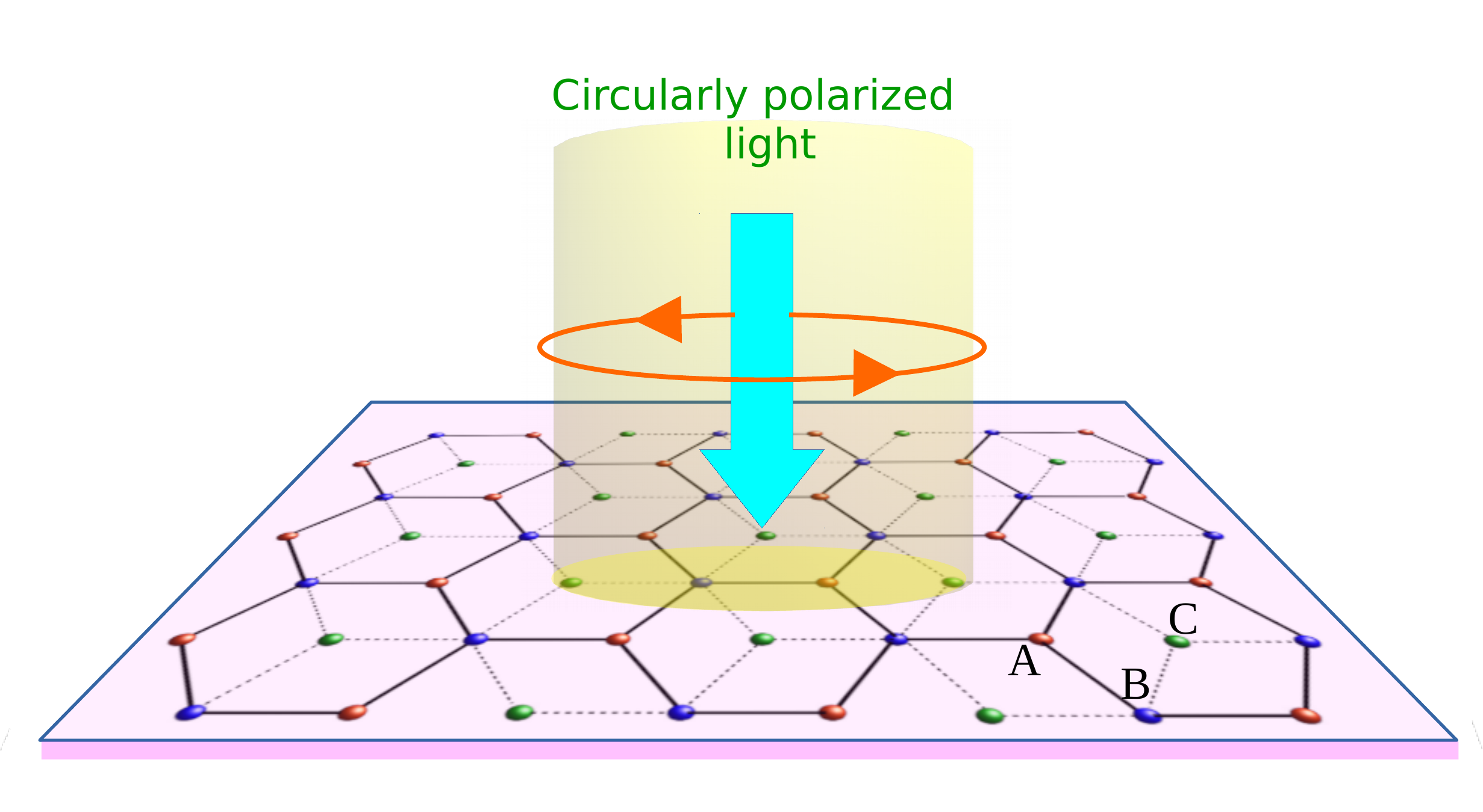}
\caption{Schematic diagram of the $\alpha$-$\mathcal{T}_3$ lattice illuminated
by off-resonant circularly polarized light.}
\label{alphat3}
\end{figure}

The $\alpha$-$\mathcal{T}_3$ lattice, as shown in Fig. \ref{alphat3}, is the extension 
of a honeycomb lattice. This is a conventional 
honeycomb lattice with two lattice points (A,B) and an additional lattice
point (C) at the centre of each honeycomb cell. 
The quasiparticle can hop from C sites to the alternate vertices (say, B) of 
the same honeycomb lattice. The hopping amplitude between A and B sites is 
$\tau \cos \phi$ and that between the B and C sites is $\tau \sin \phi$, 
where the angle $\phi$ parameterizes the hopping amplitude. 
It is convenient to express the angle $\phi$ by another parameter $\alpha $ such
that $\alpha = \tan \phi$. 
For $\phi=0$ ($\alpha = 0$), the C site is decoupled
from the honeycomb lattice and it resembles to the monolayer graphene.
The upper-left $2\times 2$ matrix block in Eq. (\ref{ham}) describes the quasiparticle 
dynamics of monolayer graphene. 
For $\alpha =1$ ($\phi =\pi/4$), the $\alpha$-$\mathcal{T}_3$ model
becomes conventional dice or $\mathcal{T}_3$ lattice having pseudospin-1 
\cite{Sutherland,Vidal,Korshunov,Rizzi,Wolf,Urban,JDMalcolm,Vigh}.
The $\alpha$-$\mathcal{T}_3$ lattice with non-zero $\alpha$ has three energy bands
since it has three sublattices consisting of a hub site (B) connected to
six rim sites (A, C). 
The dice lattice can naturally be built by growing trilayers
of cubic lattices (e.g. SrTiO${}_3$/SrIrO${}_3$/SrTiO${}_3$) in (111) 
direction \cite{Ran}.
It has been proposed theoretically that a dice lattice can be 
generated by interfering three counter-propagating pairs of identical
laser beams \cite{Rizzi} on a plane with wavelength $\lambda=3a/2$, 
with $a$ being the lattice constant.
Later, it was shown that the $\alpha$-$\mathcal{T}_3$ optical lattice can be 
achieved by dephasing one of the pairs of the laser beams while 
keeping other parameters unaltered \cite{Rizzi,alpha-t3}. 
Therefore, a continuous change of $\alpha$, through tuning phase
of one of the three pairs of the laser beams, will allow us to study the
changes in the topological properties of the system. The Hamiltonian of 
Hg${}_{1-x}$Cd${}_x$Te quantum well
can be mapped to that of low-energy $\alpha$-$\mathcal{T}_3$ model with 
effective $\alpha=1/\sqrt{3}$ on appropriate doping \cite{Malcolm}.

In recent years, there have been several studies 
\cite{alpha-t3,Illes,Tutul,Cserti,Firoz,Tutul1,Bashab,Huang,Oriekhov,Oriekhov1,Ma,Firoz1} 
on various properties of the  $\alpha$-$\mathcal{T}_3$ lattice. 
The role of variable Berry phase in orbital susceptibility  \cite{alpha-t3}, 
magnetotransport coefficients \cite{Tutul} (quantized Hall conductivity and SdH oscillation) 
and optical conductivity \cite{Illes} of the $\alpha$-$\mathcal{T}_3$ lattice has been established.
Very recently, Floquet states and the variable Berry phase dependent 
photoinduced gap in $\alpha$-$\mathcal{T}_3$ lattice irradiated by the circularly
polarized on-resonant light have been studied in detail \cite{Bashab}.   
It has been shown that an off-resonant radiation induces a gap
in graphene, on the surface states of TI \cite{TI-periodic6}, 
silicene \cite{Ezawa}, semi-Dirac systems \cite{Kush},
MoS${}_2$ \cite{Tahir} etc and transforms them to Chern insulating states. It is to be noted that the multiple Floquet bands formed by on-resonant light cannot be treated as a new static/effective band structure to determine the transport properties of such a non-equilibrium system, as shown by Kitagawa et.al.\cite{TI-periodic4}. A non-zero Chern number of any of these Floquet bands would not simply imply a quantized Hall conductance as there are additional contributions from photon-assisted electron conductions. But, on off-resonant (high-frequency) driving, the electrons cannot
directly absorb or emit any photon; only its static band structure gets modified through virtual photon absorption-emission processes. So, the off-resonant condition provides an advantage to deal with systems strongly driven out of equilibrium and the transport properties of the system may be well approximated as those originating from the effective static band structure.

Since the proposal of Haldane's Chern insulator on a honeycomb lattice \cite{FDMH},
several multiband CIs such as kagome\cite{CI-kagome1,CI-kagome2,CI-kagome4}, 
dice \cite{CI-dice1-highC} and Lieb \cite{CI-lieb} lattices 
with tunable parameters controlling the band topology have been studied.
The $\alpha$-$\mathcal{T}_3$ lattice is another example of a multiband system having
trivial topology. 
In this work, we will show that an application of circularly polarized radiation 
on the $\alpha$-$\mathcal{T}_3$ lattice makes it Haldane-type CI having non-zero
Chern number and tuning the parameter $\alpha$ leads to a topological 
phase transitions at $\alpha=1/\sqrt{2}$ by changing the Chern number
of the valence (conduction) band from $\mathcal{C}=1 (-1)$ to larger Chern number 
$\mathcal{C}=2 (-2)$. 
This phase transition results in doubling of the number of chiral 
edge modes from one to two in irradiated $\alpha$-$\mathcal{T}_3$ nanoribbon.
First we derive Floquet-Magnus Hamiltonian of the $\alpha$-$\mathcal{T}_3$ lattice 
for the entire Brillouin zone. 
We get exact analytical expressions of quasienergy band
structure over the full Brillouin zone. The triple-point degeneracy
at the Dirac points is completely removed by breaking time-reversal symmetry 
due to time-periodic circularly polarized light. 
An intriguing state, independent of the radiation amplitude, appears at 
$\alpha=1/\sqrt{2}$, where the gap between flat and valence bands 
closes at ${\bf K}$, while that between conduction and flat bands 
closes at  ${\bf K}^{\prime}$. The low-energy bands around both ${\bf K}$ and 
${\bf K}^{\prime}$ points display a Dirac-like dispersion with 
the reduced slope, as compared to monolayer graphene, at the gap closing points.

This paper is arranged as follows. In Sec. II, we provide
topological band structure of the $\alpha$-$\mathcal{T}_3$ lattice irradiated 
by the circularly polarized light. In Sec. III, we present the analytical 
calculations of the Chern number and show that the system undergoes a 
topological phase transition at $\alpha=1/\sqrt{2}$. We present 
results of chiral edge states of irradiated $\alpha$-$\mathcal{T}_3$ nanoribbon 
in Sec. IV. In Sec. V, a summary and conclusion of our results are presented.

\section{Topological band structure of $\alpha$-$\mathcal{T}_3$ lattice irradiated 
by circularly polarized light}
Considering only the nearest-neighbour (NN) hopping integrals, 
the rescaled tight-binding Hamiltonian for the $\alpha$-$\mathcal{T}_3$ lattice 
is given by
\begin{equation}\label{ham}
H_0(\textbf{k}) = \left(\begin{array}
{ccc}
0 & h(\textbf{k})\cos\phi & 0 \\  
h^*(\textbf{k})\cos\phi & 0 & h(\textbf{k}) \sin\phi \\ 
0 & h^*(\textbf{k})\sin\phi & 0
\end{array}\right),
\end{equation}
where ${\bf k} = (k_x, k_y)$ and 
$h(\textbf{k}) = \tau (\e^{i \bf{k} \cdot {\bs \delta}_1}
+ \e^{i \bf{k} \cdot {\bs \delta}_2} + \e^{i \bf{k} \cdot {\bs \delta}_3})$. 
Also, the three nearest neighbor position vectors with respect to the rim site B
are ${\bs \delta_1}= a(\sqrt{3}/2,-1/2), {\bs \delta_2} = a(-\sqrt{3}/2,-1/2)$ 
and ${\bs \delta_3} = a(0,1)$, with $a$ is the lattice constant of graphene.
The energy-wavevector dispersion, independent of $\alpha$, over the full Brillouin 
zone consists of three bands: two dispersive bands 
$E_{\pm}({\bf k}) = \pm |h({\bf k})| $ having electron-hole symmetry 
and a zero energy non-dispersive band $ E_0({\bf k}) =0$. The dispersion 
$E_{\pm}({\bf k})$ is identical to that of graphene.
The corresponding normalized eigenvectors over the full Brillouin zone are
given by 
\begin{equation}
\psi_{\textbf{k},\pm} = \frac{1}{\sqrt{2}}\left(\begin{array}
{ccc}
\cos\phi e^{-i \psi(\textbf{k})}\\
\pm 1 \\
\sin\phi e^{i \psi(\textbf{k}) }\end{array}\right), 
\psi_{\textbf{k},0}=\left(\begin{array}
{ccc}
\sin\phi e^{-i \psi(\textbf{k})}\\
0\\
-\cos\phi e^{i \psi(\textbf{k})} \end{array}\right), \nn
\end{equation}
where  $h({\bf k}) =  |h({\bf k})| e^{-i \psi({\bf k})} $.
The low-energy quasiparticles in the $\alpha$-$\mathcal{T}_3$ lattice are described 
by two-dimensional (2D) Dirac-Weyl equation. 
It is a semimetalic system in absence of any external fields/perturbations.
It will behave like a TI if a Haldane-type energy gap is induced at the Dirac points 
by external means. Next we show that circularly polarized off-resonant radiation
opens up gaps and induces topological states in $\alpha$-$\mathcal{T}_3$ lattice. 

An $\alpha$-$\mathcal{T}_3$ lattice is irradiated with circularly polarized radiation falling 
normal to the lattice plane. The corresponding vector potential
is ${\bf A}(t) = A_0 (\cos\omega t, \sin \omega t )$, where
$A_0=E_0/\omega$ with $E_0$ and $\omega$ being the electric field amplitude and
frequency of the radiation, respectively.  
By Pierl's substitution 
$ {\bf k} \rightarrow ({\bf k} + e {\bf A}(t)/\hbar)$ in Eq. \ref{ham}, we obtain
\begin{eqnarray}
H(\textbf{k},t) = \left(\begin{array}
{ccc}
0 & h(\textbf{k},t)\cos\phi & 0 \\  
h^*(\textbf{k},t)\cos\phi & 0 & h(\textbf{k},t) \sin\phi \\ 
0 & h^*(\textbf{k},t)\sin\phi & 0
\end{array}\right)
\end{eqnarray}
where $ h({\bf k},t) = \tau \sum \limits_{j=1}^{3} 
\e^{i ({\bf k}+e {\bf A}(t)/\hbar) \cdot {\bs \delta}_j} $.
The Hamiltonian $ H(\textbf{k},t) $ is periodic in time since 
${\bf A}(t+T) = {\bf A}(t)$ with the periodicity $T=2\pi/\omega$.
Using the NN vectors ${\bs \delta}_{j}$ and the Jacobi-Anger expansion 
$\e^{iz\sin\theta} = \sum \limits_{n=-\infty}^{\infty}J_n(z)\e^{i n \theta}$ with
$J_n(z)$ being the $n$-th order cylindrical Bessel function, we get
\begin{equation}
\begin{aligned}
h({\bf k},t)  & = \tau \sum \limits_{n = - \infty}^{\infty}J_n(\eta) \Big[ 
\e^{i n \omega t}  \e^{i{\bf k} \cdot {\bs \delta}_3}  +
\e^{-i n(\pi/3 + \omega t)} \e^{i{\bf k} \cdot {\bs \delta}_2} \nonumber \\ 
& + \e^{i n(\pi/3 - \omega t)} \e^{i{\bf k} \cdot {\bs \delta}_1} \Big].
\end{aligned}
\end{equation}
Here $ \eta = eA_0a/\hbar$ is a dimensionless parameter characterizing 
the light intensity (can be expressed as square root of intensity and fine structure constant).
Typically, $\eta \ll 1$ for the intensity of lasers and pulses available
in the THz frequency domain. 
The off-resonant condition can be achieved when the photon energy is much larger than
the band width of the undriven system i. e. $\hbar \omega  > 6\tau$.
When the light frequency satisfies the off-resonant condition, the band
structure modifies by the second-order virtual photon absorption-emission 
processes.

The effective time-independent Hamiltonian valid under off-resonant 
condition \cite{Magnus,Magnus-review,Magnus-Schlieman} is 
\begin{equation}\label{eff}
H_{\rm eff}(\textbf{k}) = H_0(\textbf{k}) + [H_-({\bf k}),H_+({\bf k})]/\hbar\omega
+ \mathcal{O}(1/\omega^2),
\end{equation}
where
$$
H_{\pm}({\bf k}) = \frac{1}{T} \int_0^T dt \; e^{\mp i \omega t} H({\bf k}, t)
$$
is the Fourier component of the Hamiltonian $H({\bf k}, t)$.
By Fourier transform, we obtain
\begin{eqnarray}
[H_-({\bf k}),H_+({\bf k})] = 
\frac{\Delta({\bf k})}{2}S_z(\alpha)
\end{eqnarray}
where $S_z(\alpha)$ is defined as
\begin{eqnarray}
S_z(\alpha)= 2\left(\begin{array}
{ccc}
\cos^2\phi & 0 & 0 \\  
0 & -\cos 2\phi & 0 \\ 
0 & 0 & -\sin^2\phi
\end{array}\right)
\end{eqnarray}
and $ \Delta({\bf k}) = |g({\bf k})|^2-|f({\bf k})|^2 = \hbar\omega \; \gamma({\bf k}) $
with 
\begin{eqnarray}
g({\bf k}) & = & \tau J_1 (\eta) 
\big[\e^{i {\bf k} \cdot {\bs \delta}_1} \e^{i\pi/3} + 
\e^{i {\bf k} \cdot {\bs \delta}_2} \e^{-i\pi/3} - 
\e^{i {\bf k} \cdot {\bs \delta}_3}  \big] \\
f({\bf k}) & = & \tau J_1 (\eta) 
\big[-\e^{i {\bf k} \cdot {\bs \delta}_1} \e^{-i\pi/3} - 
\e^{i {\bf k} \cdot {\bs \delta}_2} \e^{i\pi/3} + 
\e^{i {\bf k} \cdot {\bs \delta}_3}  \big].
\end{eqnarray}
Hence, the light-matter coupling results in a mass term of the form 
$\gamma({\bf k})S_z(\alpha)/2$ which lifts the three-fold degeneracy at 
the Dirac points. It can be shown that the mass term reduces to 
$\mu\beta^2\hbar\omega S_z(\alpha)/2$ in the linearized low energy limit 
where $\beta=3\eta \tau/(2\hbar\omega)$ and $\mu=1(-1)$ corresponds to 
${\bf K}({\bf K^\prime})$ valleys. On time-reversal operation, 
$\gamma({\bf k})$ changes sign, which implies the breaking of TRS 
in the system. Similarly, the term also changes sign on switching 
from right to left circular polarization. So, the mass term is trivially 
zero in case of linearly polarized light since it is a linear 
combination of both the polarizations with equal weights. 

Interestingly, the mass term is Haldane-type which has opposite signs 
in the two valleys. It has been shown that the effective Hamiltonian in irradiated graphene under off-resonant consition can be mapped to the Haldane model with no sublatice potential and complex next nearest neighbour hoppings, which break TRS \cite{TI-periodic4}. In Haldane model, the NN hoppings do not accumulate 
Aharonov-Bohm (AB) phases since the net magnetic flux through a unit cell 
is zero. In this model, the magnetic flux is locally zero everywhere 
on the lattice plane at a given time. But the time-varying vector potential 
is spatially constant over the lattice, due to which NN hoppings acquire 
time-dependent AB phases.

The effective Hamiltonian (\ref{eff}) can now be written explicitly as 
\begin{equation}\label{heff}
H_{\rm eff}(\textbf{k}) = \left(\begin{array}
{ccc}
\gamma({\bf k})\cos^2\phi & h(\textbf{k})\cos\phi & 0 \\  
h^*(\textbf{k})\cos\phi & -\gamma({\bf k})\cos 2\phi & h(\textbf{k}) \sin\phi \\ 
0 & h^*(\textbf{k})\sin\phi & -\gamma({\bf k})\sin^2\phi
\end{array}\right).
\end{equation}
The Hamiltonian $H_{\rm eff}({\bf k})$ satisfies the following anti-commutation
relations for $\alpha=0 $ and $\alpha=1$: 
\begin{equation} \label{anti-com}
\{ H_{\rm eff}^G({\bf k}),\mathcal{P}_G \}=0, \hspace{0.3cm} 
\{H_{\rm eff}^D({\bf k}),\mathcal{P}_D \}=0.
\end{equation}
Here $\mathcal{P}_G$ and $\mathcal{P}_D$ are operators 
defined for graphene and dice respectively as follows:
\begin{equation}
\mathcal{P}_G=\mathcal{K}\left(\begin{array}{ccc}
0 &-1 & 0\\
1 & 0 & 0\\
0 & 0 & 1
\end{array}\right), 
\hspace{0.3cm}
\mathcal{P}_D=\mathcal{K}\left(\begin{array}{ccc}
0&0&-1\\
0&1&0\\
-1&0&0
\end{array}\right)
\end{equation}
with $\mathcal{K}$ being the complex-conjugation operator. 
The relations (\ref{anti-com}) imply that a band with energy
$\epsilon({\bf k})$ will have a partner band with energy $-\epsilon({\bf k})$.
This symmetry confirms the presence of a
zero energy band in the three-band system.
Hence, the flat band in 
dice lattice will not be perturbed by radiation. Also, the Hamiltonian is 
traceless, implying the sum of energies of the bands will be zero 
for all values of $\alpha$.

The eigen values $\epsilon_m(\textbf{k})$ of $H_{\rm eff}(\textbf{k})$ represent 
the off-resonant quasienergy band structure. 
The characteristic equation for the eigen value problem turns out to 
be a $depressed$ cubic equation:
$ \lambda^3 + p \lambda + q = 0 $,
where
\begin{eqnarray}
p & = & - \Big[ |h({\bf k})|^2 + \gamma({\bf k})^2 
\big(\cos^2 2\phi + \frac{\sin^2 2\phi}{4}\big) \Big] \\
q & = & - \frac{\gamma({\bf k})^3}{4}\sin^2 2\phi \cos2\phi.
\end{eqnarray}
The eigen values are of the form
\begin{equation}
\epsilon_m({\bf k}) = 2\sqrt{\frac{-p}{3}} \cos\Big[ \frac{1}{3} 
\cos^{-1}\bigg(\frac{3q}{2p}\sqrt{\frac{-3}{p}}\bigg)-\frac{2\pi m}{3} \Big]
\end{equation}
with $m$=0,1 and 2 correspond to the quasienergies of 
the conduction, flat and valence bands, respectively.
The band structure of the undriven $\alpha$-$\mathcal{T}_3$ lattice is strongly modified 
by the off-resonant radiation and becomes $\alpha$ as well as $\eta$ dependent. 
It exhibits interesting features as we tune $\alpha$, which will be discussed in detail. 
The components of the normalized eigen vectors $| \Psi_m({\bf k}) \rangle  =
\big(a_m({\bf k})\hspace{.2cm}b_m({\bf k})\hspace{.2cm}c_m({\bf k})\big)^T$
can be written as
\begin{eqnarray}\label{eigenveca}
a_m({\bf k}) & = & \frac{d({\bf k}) \sin\theta\cos\phi 
\e^{-i\psi({\bf k})}}{\epsilon_m({\bf k}) - d({\bf k}) \cos\theta\cos^2\phi}
b_m({\bf k}) \\\label{eigenvecb}
c_m({\bf k}) & = & \frac{d({\bf k}) \sin\theta\sin\phi 
\e^{i\psi({\bf k})}}{\epsilon_m({\bf k})  +  d({\bf k}) \cos\theta\sin^2\phi}
b_m({\bf k})
\end{eqnarray}
with
\begin{eqnarray}
b_m({\bf k}) & = & \Big[1 + \Big(\frac{d({\bf k}) \sin\theta\cos\phi}{\epsilon_m({\bf k}) -
d({\bf k}) \cos\theta\cos^2\phi}\Big)^2  \nonumber \\
& + & \Big(\frac{d({\bf k}) \sin\theta\sin\phi}{\epsilon_m({\bf k}) + 
d({\bf k}) \cos\theta\sin^2\phi}\Big)^2 \Big]^{-1/2},
\end{eqnarray}
where we have parameterized $\gamma({\bf k})$ and $h({\bf k})$ as
$\gamma({\bf k}) = d({\bf k}) \cos\theta$, 
$h({\bf k}) = d({\bf k}) \sin\theta \e^{-i\psi({\bf k})}$,
with $d({\bf k}) = \sqrt{|h({\bf k})|^2 + \gamma({\bf k})^2} $.  

\begin{widetext}
\begin{figure*}[htbp]
\includegraphics[trim={3cm 0cm 0cm  0cm},clip,width=16cm]{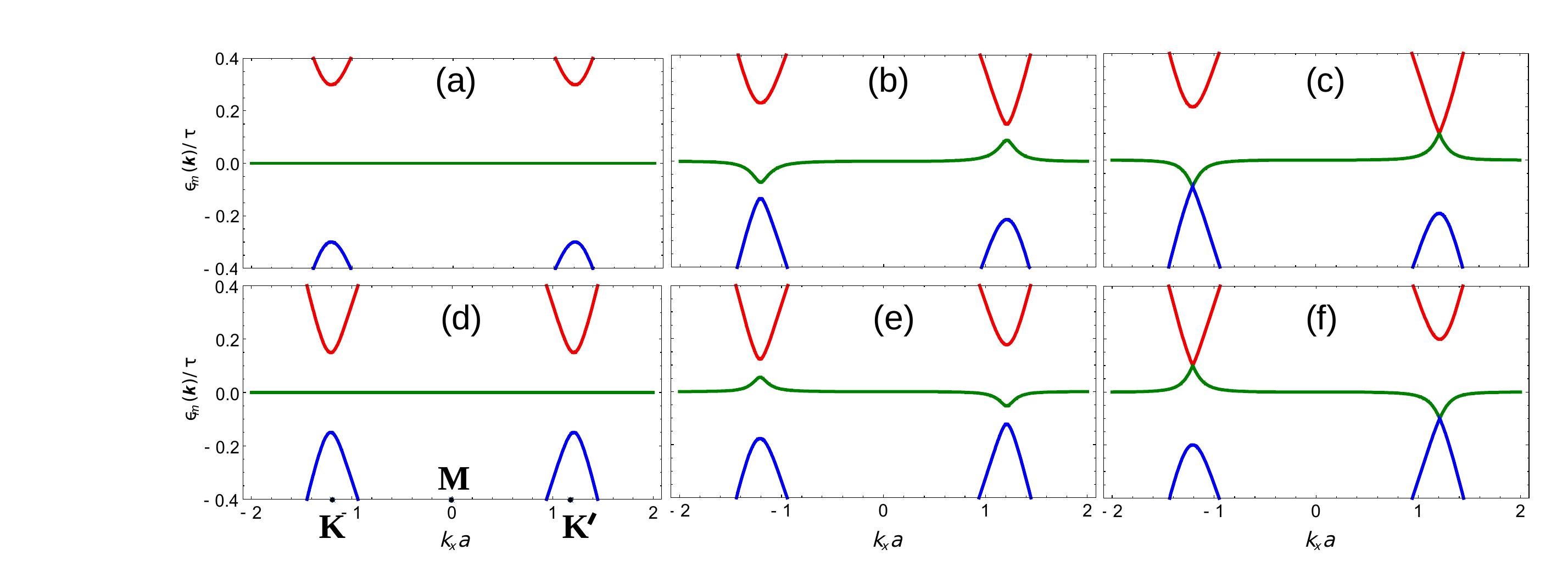}
\caption{Plots of low-energy Floquet bands for various values of $\alpha$: 
(a) $\alpha=0$, (b) $\alpha = 0.6$, (c) $\alpha = 1/\sqrt{2}$, (d) $\alpha=1$, 
(e) $\alpha=1.2$ and (f) $\alpha = \sqrt{2}$. The bands are plotted along 
the line joining the high-symmetry ${\bf K, M }$ and ${\bf K^\prime}$ points. 
Bands are no longer symmetric under exchange of valleys except at $\alpha=0$ and 
$\alpha=1$. Here we have taken $J_1(\eta) = 0.57$.}
\label{floquet-plot-section}
\end{figure*}
\end{widetext}

\begin{figure}[htbp]
\includegraphics[trim={0cm 0cm 0cm  0cm},clip,width=8cm]{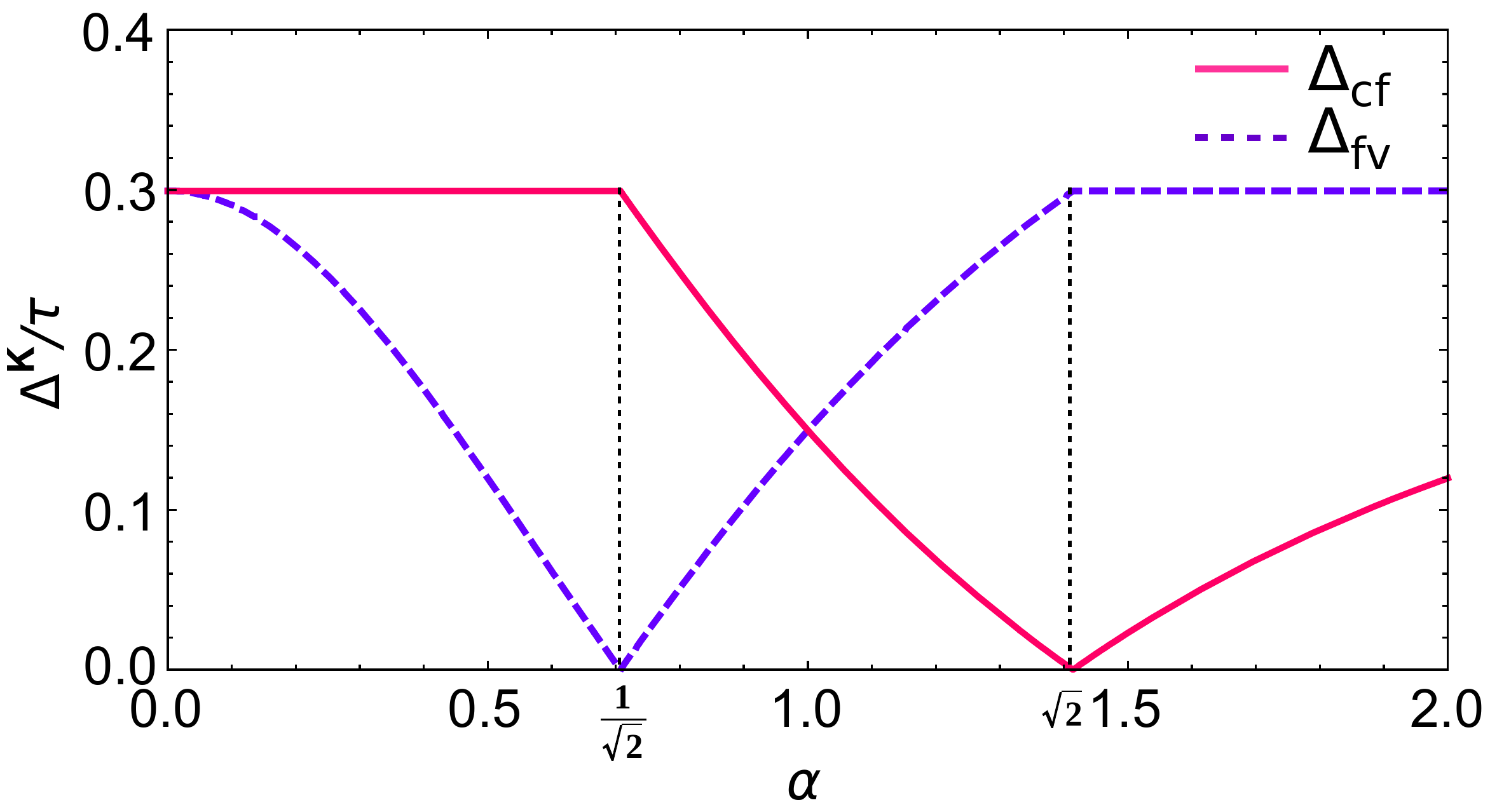}
\caption{Plots of the photoinduced gaps at the Dirac point ${\bf K}$ as a function of $\alpha$. }
\label{gaps-dirac}
\end{figure}

Now we shall analyze the topological band structures as we vary 
$\alpha$ continuously. Figure \ref{floquet-plot-section} shows the low-energy topological 
bands for different values of $\alpha$ in the first Brillouin zone. 
For three-band systems, there can be two distinct band gaps at the Dirac points:
gap between i) conduction and flat bands ($\Delta_{\rm cf}^{ {\bf K}/{\bf K}^{\prime} }$); 
and ii) flat and valence bands  ($\Delta_{\rm fv}^{ {\bf K}/{\bf K}^{\prime} }$) at 
${\bf K}/{\bf K}^{\prime}$ points. 
In presence of TRS 
breaking circularly polarized light, the triple point degeneracy at both the 
Dirac points is completely lifted 
(i.e. $\Delta_{\rm cf}^{ {\bf K}/{\bf K}^{\prime} } \neq 0$ and 
$\Delta_{\rm fv}^{ {\bf K}/{\bf K}^{\prime} } \neq 0$) except at
$\alpha = 1/\sqrt{2}$.
The photoinduced gaps at $\alpha=0$ and 1 are 
$\Delta_{\rm cf}^{ {\bf K}/{\bf K}^{\prime} } =
\Delta_{\rm fv}^{ {\bf K}/{\bf K}^{\prime} } = \beta^2\hbar \omega $
and $\Delta_{\rm cf}^{ {\bf K}/{\bf K}^{\prime} } = 
\Delta_{\rm fv}^{ {\bf K}/{\bf K}^{\prime} } =  \beta^2\hbar \omega/2$, 
respectively.
It is interesting to note that
$\Delta_{\rm cf}^{\bf K} = 0$ but 
$\Delta_{\rm fv}^{{\bf K}} \neq  0$  
and $\Delta_{\rm cf}^{{\bf K}^{\prime} } \neq 0$ but
$\Delta_{\rm fv}^{{\bf K}^{\prime}} =  0$ 
at $\alpha=1/\sqrt{2}$.
It implies that the band gaps at the Dirac points do not open completely
at $\alpha=1/\sqrt{2}$.   
Note that this result is independent of the radiation amplitude $\eta$ 
(as long as off-resonant approximation is valid).
The partial closing of the band gap at $\alpha = 1/\sqrt{2}$ 
can be deduced by obtaining the eigen values at a Dirac 
point (say \textbf{K}) viz. $\epsilon_0 = \cos^2\phi$, 
$\epsilon_1 = - \cos2\phi$, $\epsilon_2 = - \sin^2\phi$.
Equating $\epsilon_0$ with $\epsilon_1$, we find that the band touching 
occurs at $\alpha = 1/\sqrt{2}$. We present plots of $ \Delta_{\rm cf}^{\bf K} $ and
$\Delta_{\rm fv}^{ {\bf K} }$ vs $\alpha$ in Fig. \ref{gaps-dirac}. 
The system exhibits an interesting property of $\alpha \rightarrow 1/\alpha$ duality. 
The measurable quantities of the system will be same for $\alpha$ and $1/\alpha$. Hence, 
similar gap-closing is also seen at $\alpha=\sqrt{2}$. However, the 
duality exchanges the Dirac points i.e. 
$ \Delta_{\rm cf}^{\bf K} (\alpha)= \Delta_{\rm cf}^{\bf K^\prime} (1/\alpha) $ 
and $ \Delta_{\rm fv}^{\bf K} (\alpha) = \Delta_{\rm fv}^{\bf K^\prime} (1/\alpha) $. 
The band gaps (in units of $\beta^2\hbar \omega$) at the 
Dirac point ${\bf K}$  are tabulated in table I.
For ${\bf K}^{\prime}$ point, one can easily check that
$\Delta_{\rm cf}^{ {\bf K}^{\prime} } = \Delta_{\rm fv}^{ {\bf K} } $ and
$\Delta_{\rm fv}^{ {\bf K}^{\prime} } = \Delta_{\rm cf}^{ {\bf K} } $ for given $\alpha$.
\begin{table}[H]
\begin{center}
\begin{tabular}{ | c | c | c | c|}
\hline
Gaps & 0 $ \leq \alpha \leq 1/\sqrt{2}$ & $1/\sqrt{2} \leq \alpha \leq \sqrt{2}$ & $ \alpha \geq \sqrt{2}$   \\
\hline
$ \Delta_{\rm cf}^{\bf K} $ & 1 & $ \cos^2\phi + \cos 2\phi$  & $-(\cos 2\phi+\cos^2\phi) $   \\
\hline
$ \Delta_{\rm fv}^{\bf K} $  &  $ -\sin^2\phi + \cos 2\phi$ &  $ \sin^2\phi - \cos 2\phi$  & 1    \\
\hline
\end{tabular}
\caption{Photoinduced gaps at the Dirac point ${\bf K}$ as a function of $\alpha$.}
\end{center}
\end{table}

Substituting ${\bf k} = {\bf K} + {\bf q}$ with ${\bf q} \rightarrow 0$ in 
Eq. (\ref{heff}), we get low-energy Hamiltonian around ${\bf K}$. 
Interestingly, the touching bands i.e. flat and conduction bands exhibit 
Dirac cones in the low-energy limit as
$$
\epsilon_{0,1}({\bf q}) = \frac{\beta^2 \hbar \omega}{3} \pm 
\frac{\hbar v_f}{\sqrt{3}} |{\bf q}|.
$$ 

It is to be noted that in the field free case, 
$\epsilon_m({\bf k}) = \epsilon_m(-{\bf k})$ 
for all values of $\alpha$. But for the irradiated model,  
we have $\epsilon_m({\bf k}) = \epsilon_m({-\bf k})$ for $\alpha=0,1$, 
and $\epsilon_m(-{\bf k})\neq \epsilon_m({\bf k})$ for $\alpha \neq 0, 1$. 
This can be explained as follows $-$ In the radiation-free case, 
the Kramer's degeneracy ensured by TRS guarantees 
$\epsilon_m({\bf k}) = \epsilon_m(-{\bf k})$ irrespective of the value 
of $\alpha$ or other symmetries.  On application of TRS-breaking circularly 
polarized light, the Kramer's degeneracy is lifted. Now, the band structure 
will be an even function in ${\bf k}$ only if the lattice has inversion symmetry (IS). 
Since  graphene and dice lattice have IS, the band structure remains an 
even function in quasimomentum despite a broken TRS. 

The topology of the band structure remains same if the energy gap
in the band structure does not close and reopen while tuning 
the parameter continuously.
Here we have seen that one of the gaps closes at $\alpha=1/\sqrt{2}$ and 
reopens when $  1/\sqrt{2} < \alpha < \sqrt{2}$. Hence, we expect a transition 
in band topology at $\alpha=1/\sqrt{2}$. 
In 2D, a change in the Chern number or TKNN integer can be 
used to identify whether the system undergoes a topological
transition or not. In the next section, we show that there is indeed a 
topological phase transition at  $\alpha=1/\sqrt{2}$ 
(equivalently at $\alpha=\sqrt{2}$) by evaluating the
Chern number explicitly as a function of $\alpha$.

\section{Calculations of Chern number and topological phase transitions}
We need to analyze the behaviour of the Berry connection and 
Berry curvature in order to calculate the Chern number of each 
band analytically \cite{kohmoto,analytic,analytic1} as well as numerically 
\cite{numerical}.
The Berry connection for the band $\epsilon_m({\bf k})$ can be written as
$\textbf{A}_m({\bf k}) = i \langle \Psi_m({\bf k})| {\bs \nabla}_\textbf{k}| 
\Psi_m({\bf k})\rangle$.
Under the gauge used in (\ref{eigenveca}) and (\ref{eigenvecb}), the Berry connection reduces to
\begin{eqnarray} \label{connection}
\textbf{A}_m({\bf k}) 
= s_m({\bf k}){\bs \nabla}_{{\bf k}}\psi({\bf k}),
\end{eqnarray}
with $s_m({\bf k}) = [|a_m({\bf k})|^2 - |c_m({\bf k})|^2] $. 
The Berry curvature of the $m$-th band is defined as
\begin{eqnarray} \label{curvature}
\Omega_m({\bf k}) = \hat {\bf z} \cdot 
[{\bs \nabla}_{\bf k} \times \textbf{A}_m({\bf k}) ].
\end{eqnarray}
\begin{figure}[htbp]
\hspace{-.5cm}	\includegraphics[trim={0cm 0cm 0cm  0cm},clip,width=9cm]{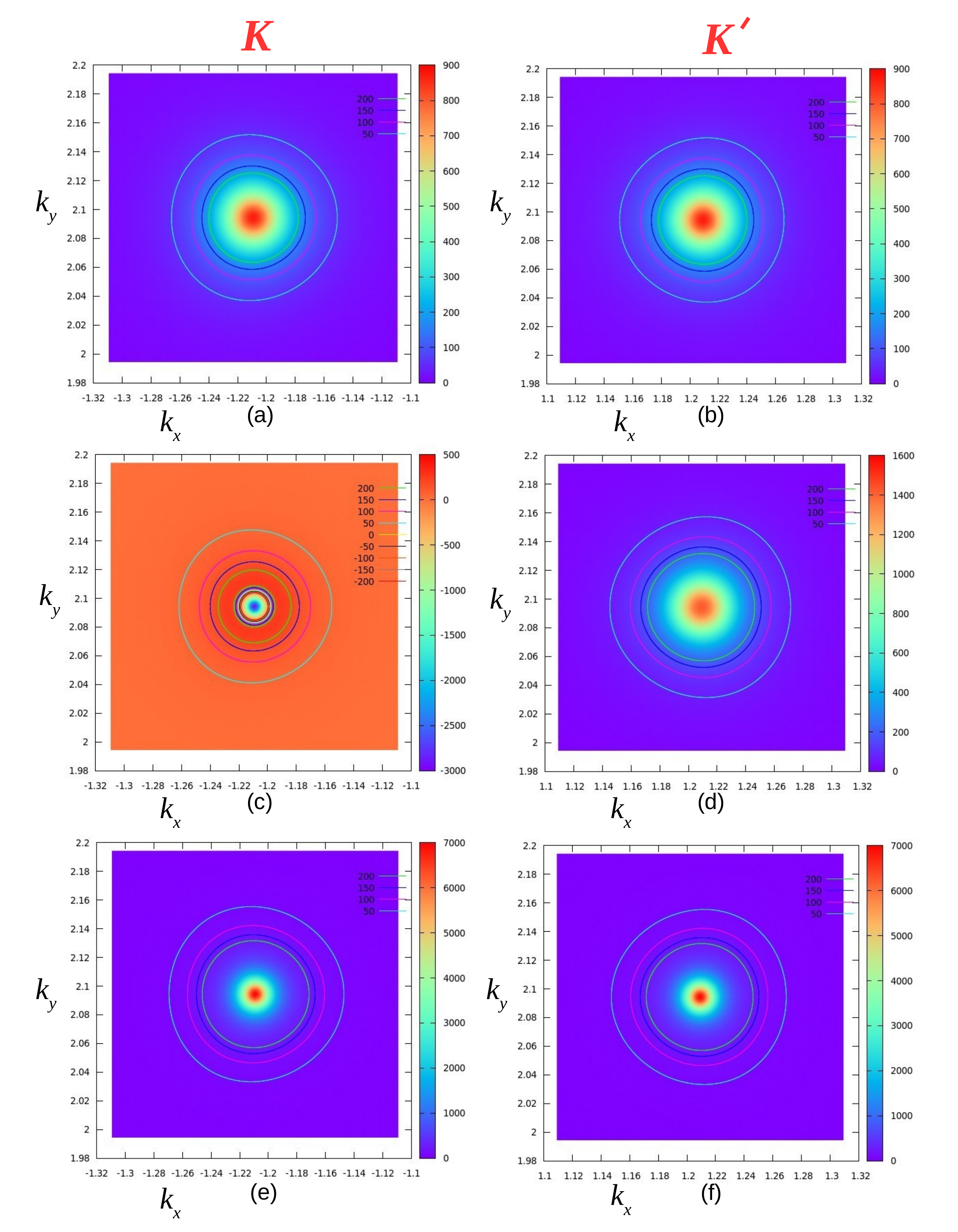}
\caption{Density-contour plots of the Berry curvature ($\Omega({\bf k})$) around 
the two Dirac points ${\bf K}$ and ${\bf K}^{\prime}$ for different values of $\alpha$: 
Top panel: $\alpha=0$,  Middle panel: $\alpha=0.5$, Bottom panel: $\alpha=1$. 
Here $k_x$ and $k_y$ are in units of $a^{-1}$.}
\label{berry-plot}
\end{figure}
Figure (\ref{berry-plot}) shows the plots of $\Omega_2({\bf k})$ around 
the two valleys for three values of $\alpha$. The distribution of 
$\Omega_2({\bf k})$ in graphene [Fig. \ref{berry-plot}-(a),(b))] is 
identical in the two valleys. Since ${\bf K}={-{\bf K}^\prime}$, $\Omega_2({\bf k})$ 
is an even function indicating the presence of inversion (IS) symmetry 
in the lattice, but a broken TRS. This also holds true for $\alpha=1$ i.e. dice lattice 
[Fig. \ref{berry-plot}-(e),(f))]. However, for $\alpha=0.5$,  $\Omega_2({\bf k})$ 
is largely different in the two valleys [Fig. \ref{berry-plot}-(d), (e))]. 
This is a signature of the absence of IS and a broken TRS.

The Berry connection depends on the gauge and may have singularities 
within the first Brillouin zone (FBZ).
The Berry curvature of the $m$-th band is well defined when the 
quasienergy $\epsilon_m({\bf k})$ is non-degenerate (i.e. $m$-th band
does not touch any other bands) for all values of ${\bf k}$ within the
FBZ. Contour plots of the Berry curvature near the two Dirac points
${\bf K}$ and ${\bf K}^{\prime}$ for different values of $\alpha$ are 
shown in Fig. \ref{berry-plot}.
The surface integral of Berry curvature $\Omega_m({\bf k})$ over the 
FBZ gives $2 \pi \mathcal{C}_m$, where $\mathcal{C}_m$ is an integer 
called the Chern number or TKNN index \cite{tknn} for the $m$-th band:
\begin{eqnarray} \label{Cnumber}
\mathcal{C}_m = \frac{1}{2\pi} \int_{\rm FBZ} \hat {\bf z} \cdot
[{\bs \nabla}_{\bf k} \times \textbf{A}_m({\bf k}) ] \; d^2{\bf k}.
\end{eqnarray}
It is important to mention here that any contributions due to gauge-dependent singularities 
in $\textbf{A}_m({\bf k})$ must be excluded from the above equation.
If a global gauge transformation removes all the singularities, then Chern number 
of the band will be trivially zero by Stokes theorem.
Note that $\mathcal{C}_m \neq 0$ implies the absence of a global gauge under which 
the Berry connection has no singularities in the FBZ.
The Berry connection 
${\bf A}_m({\bf k})$ given in Eq. (\ref{connection}) is proportional to 
$\nabla_{\bf k}\psi({\bf k})$.
The gauge-dependent singularities in the Berry connection $\textbf{A}_m({\bf k})$
arise at the ${\bf k}$ points where the phase $\psi({\bf k})$ of the off-diagonal
matrix elements $h({\bf k}) $ is ill defined.
It occurs if the function $h({\bf k}) = 0 $ for certain values of ${\bf k}$.
In this band structure, since $h({\bf K}) = h({\bf K}^{\prime}) = 0$,
$\psi({\bf K})$ and $\psi({\bf K}^{\prime})$ are not
defined and hence there may be singularity in $\textbf{A}_m({\bf k})$ at the Dirac
points if $s_m({\bf K}) \neq 0$ or $s_m({\bf K}^{\prime}) \neq 0$.
Thus, we expect a non-zero Chern number in this case.
\begin{figure}[htbp]
\hspace{-.5cm}\includegraphics[trim={0cm 0cm 0cm  0cm},clip,width=8cm]{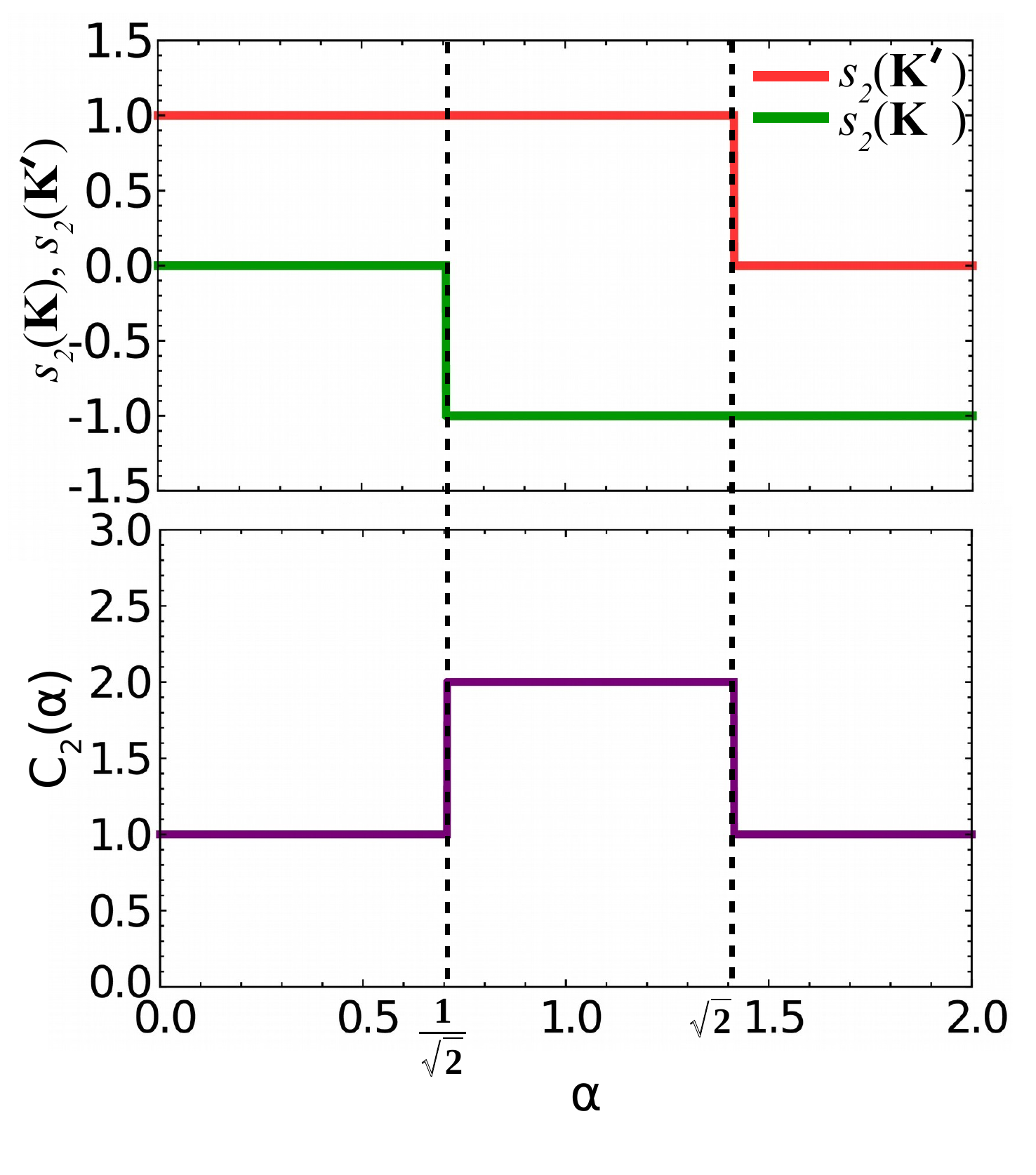}
\caption{Top panel: Plots of $s_2({\bf K})$ and $s_2({\bf K}^{\prime})$ vs $\alpha$.
Bottom panel: Plots of the Chern number $\mathcal{C}_2$ vs $\alpha$.}
\label{singularity}
\end{figure}
First we calculate the Chern number for the valence band corresponding to $m=2$.
The variation $s_2({\bf K})$ and $s_2({\bf K}^{\prime})$ with $\alpha$ is displayed
in Fig. \ref{singularity}. 
Note that $s_2({\bf K})$ and $s_2({\bf K}^{\prime})$ are
evaluated very close to the Dirac points since they are not defined exactly at
the Dirac points.  
These two functions can be written mathematically as
\begin{eqnarray}\label{sk}
s_2({\bf K}) = -\Theta(\alpha- 1/\sqrt{2}), \hspace{0.3cm}
s_2({\bf K}^{\prime}) = \Theta(\sqrt{2}-\alpha),
\end{eqnarray}
where $\Theta(x)$ is the usual unit step function.
\begin{figure}[htbp]
\includegraphics[trim={2cm 1.5cm 2cm  1.7cm},clip,width=8cm]{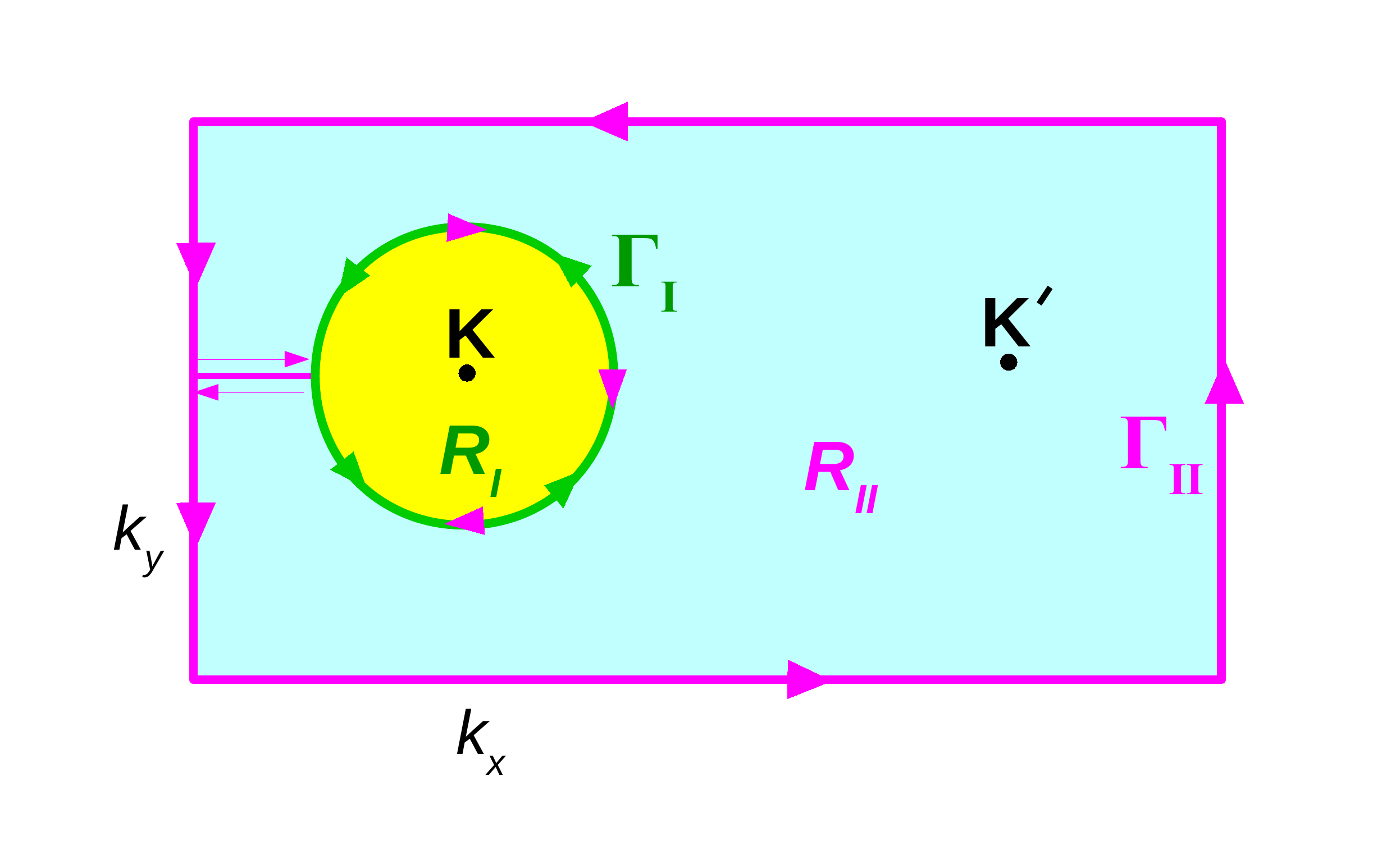}
\caption{Sketch of the FBZ with the locations of the singular points.}
\label{BZ-singularity}
\end{figure}

Now we calculate the Chern number of the valence band for $\alpha < 1/\sqrt{2} $.
Since $s_2({\bf K^\prime}) = 1$ and $s_2({\bf K}) = 0$, we have 
${\bf A}_2({\bf K}^\prime)$ not defined and  ${\bf A}_2({\bf K})=0$. 
For convenience, we remove the subscript {\bf 2} from ${\bf A}_2$, as 
we will stick to the quantities related to the valence band only.
The Berry connection for valence band under the chosen gauge 
(say ${\bf A}_{\rm I}({\bf k})$) has a singularity at ${\bf K}^{\prime}$ point. 
Hence, ${\bf A}_{\rm I}({\bf k})$ 
is not smoothly defined across the FBZ. We make a gauge 
transformation ${\bf A}_{\rm II}({\bf k}) \rightarrow {\bf A}_{\rm I}({\bf k}) - 
{\bs \nabla}_{\bf k} \psi({\bf k}) = [s_{\rm I}({\bf k}) - 1] {\bs \nabla}_{\bf k} 
\psi({\bf k})$, which gives 
${\bf A}_{\rm II}({\bf K}^\prime) = 0$ and  ${\bf A}_{\rm II}({\bf K})$ not defined. 
In this gauge, ${\bf K}$ is the singular point. As long as there is a singularity 
under a given gauge, integral of the Berry curvature will not be defined if the gauge
is chosen globally across the FBZ. So, we divide the FBZ, as depicted in 
Fig. \ref{BZ-singularity}, 
into two regions $ R_{\rm I}$ and $R_{\rm II}$ surrounding ${\bf K}$ and ${\bf K}^{\prime}$,
respectively. We assign gauge-related Berry connections ${\bf A}_{\rm I}({\bf k})$
and ${\bf A}_{\rm II}({\bf k})$ in $R_{\rm I}$ and $R_{\rm II}$, respectively,
so that the Berry curvature ($ \Omega({\bf k})$) obtained from them is
well-defined in each region. $\Gamma_{\rm I}$ and $\Gamma_{\rm II}$ are contours enclosing 
$R_{\rm I}$ and $R_{\rm II}$ respectively. The two regions share a common boundary 
coinciding with $\Gamma_{\rm I}$.
Now, the Chern number can be written as
\bearr
\mathcal{C}_2 & = & \frac{1}{2\pi}\l[\int_{R_{\rm I}} {\bs \nabla}_{\bf k} \times {\bf A}_{\rm I}({\bf k}) +
\int_{R_{\rm II}} {\bs \nabla}_{\bf k} \times {\bf A}_{\rm II}({\bf k}) \r] \cdot \hat{\bf z}
d^2 {\bf k} \nn \\
& = & \frac{1}{2\pi} \l[\oint_{\Gamma_{\rm I} } {\bs \nabla}_{\bf k} \psi({\bf k}) \cdot d{\bf k} \r]
\eearr
where we have used the fact that integral along outer boundary of $\Gamma_{\rm II}$ vanishes 
due to periodicity in ${\bf A(k)}$ across the FBZ.
The region $R_{\rm I}$ can be chosen as an infinitesimally small circle
around the Dirac point {\bf K}. Then, the term within the brackets
is the vorticity $v_{\bf K}$ around ${\bf K}$ point. Since $v_{\bf K} = 2\pi$,
$\mathcal{C}_2 = 1$.
The valence band is degenerate with flat band at $\alpha =1/\sqrt{2}$. Hence, the Chern number
of the valence band at $\alpha=1/\sqrt{2}$ is not defined.

We have already seen that the three bands are non-degenerate again when
$\alpha$ lying between $ 1/\sqrt{2}$ and $\sqrt{2}$ i.e. 
$ 1/\sqrt{2} < \alpha < \sqrt{2}$.
In this case, $s_{\rm I}({\bf K}^{\prime})= 1 $ and $s_{\rm I}({\bf K}) = - 1$.
Hence, both ${\bf A}_{\rm I}({\bf K})$ and ${\bf A}_{\rm I}({\bf K^\prime})$ are not
defined. Since, we need atleast one non-singular point, this gauge choice
is redundant. However, the gauge transformed 
${\bf A}_{\rm II}({\bf K}^{\prime}) = [s_{\rm I}({\bf K}^{\prime}) - 1] {\bs \nabla}_{\bf k} 
\psi({\bf K}^{\prime}) = s_{\rm II}({\bf K}^{\prime}) {\bs \nabla}_{\bf k} \psi({\bf K}^{\prime})=0$
and ${\bf A}_{\rm II}({\bf K)} = [s_{\rm I}({\bf K}) - 1] 
{\bs \nabla}_{\bf k} \psi({\bf K}) = s_{\rm II}({\bf K}) {\bs \nabla}_{\bf k} \psi({\bf K}) 
= -2 {\bs \nabla}_{\bf k} \psi({\bf K})$. Thus, we have 
$s_{\rm II}({\bf K}^{\prime}) = 0$ and $s_{\rm II}({\bf K}) = - 2$, i.e. 
${\bf A}({\bf k}) $ is singular at ${\bf K}$ and is non-singular at ${\bf K}^{\prime}$. 
On making a gauge transformation ${\bf A}_{\rm III}({\bf k)} \rightarrow 
{\bf A}_{\rm II}({\bf k)} + 2 {\bs \nabla}_{\bf k}(\psi ({\bf k})) = 
[s_{\rm II}({\bf k}) + 2) {\bs \nabla}_{\bf k} \psi({\bf k}) = s_{\rm III}({\bf k}) 
{\bs \nabla}_{\bf k} \psi({\bf k})$. Now, we have 
$s_{\rm III}({\bf K}^{\prime}) = 2$ and 
$s_{\rm III}({\bf K}) = 0$, i.e. ${\bf K}$ is non-singular.

Again, we divide the BZ, similar to Fig. \ref{BZ-singularity}, into 
two regions $ R_{\rm II}$ and $ R_{\rm III}$ surrounding ${\bf K}^{\prime}$ 
and ${\bf K}$, respectively. We assign gauge-related Berry connections 
${\bf A}_{\rm II}({\bf k})$ and ${\bf A}_{\rm III}({\bf k})$ in $R_{\rm II}$ 
and $R_{\rm III}$, respectively, such that that $\Omega({\bf k}$) is well-defined 
in each region. So, the Chern number is given by
\bearr
\mathcal{C}_2 & = & \frac{1}{2\pi} \l[\int_{R_{\rm II}} {\bs \nabla}_{\bf k} \times 
{\bf A}_{\rm II}({\bf k}) + \int_{R_{\rm III}} {\bs \nabla}_{\bf k} 
\times {\bf A}_{\rm III}({\bf k}) \r] 
\cdot \hat{\bf z} d^2 {\bf k} \nn \\
& = & \frac{1}{2\pi} \l[2\oint_{\Gamma_{\rm I} } {\bs \nabla}_{\bf k}\psi({\bf k}) 
\cdot d{\bf k} \r ] = 2,
\eearr
where we have taken the infinitesimal loop around ${\bf K}$ point to have
the positive sense of rotation.

The Chern number for the valence band for all $\alpha$ as shown 
in Fig. \ref{singularity} can be expressed as
\begin{eqnarray}
\mathcal{C}_2(\alpha) = \Theta(1/\sqrt{2}-\alpha) +
2\Theta(\alpha-1/\sqrt{2}) - \Theta(\alpha - \sqrt{2}). \nn
\end{eqnarray}
The Chern number for the non-degenerate flat band turns out to be zero for all values of $\alpha$
i.e. $\mathcal{C}_1(\alpha) = 0$. Therefore, the Chern number for the conduction band
corresponding to $m=0$ is $ \mathcal{C}_0( \alpha) = - \mathcal{C}_2(\alpha)$.
Using Eqs. (\ref{curvature}) and (\ref{Cnumber}), we have also calculated
the Berry curvature and Chern numbers $\mathcal{C}_m$ numerically.
Our numerical results support the exact analytical results.
Figure \ref{singularity} displays that the system undergoes a topological 
phase transition at $\alpha=1/\sqrt{2}$ (also at $\alpha = \sqrt{2}$ due 
to the duality) since there is a change in the Chern number.

The anomalous Hall conductivity is directly related to the Chern number.
When the Fermi energy is located in a band gap, the Hall conductivity can be 
expressed in terms of the Chern number as
$\sigma_{H} = (e^2/h) \sum_{m}^{} \mathcal{C}_m$,
where $m$ is restricted to the filled bands below the Fermi energy.
By complete filling of the valence band or both the valence and flat bands,
the system becomes a QHI with the Hall conductivity
$\sigma_H = e^2/h$ for $\alpha < 1/\sqrt{2}$ and
$\sigma_H = 2e^2/h$ for $1/\sqrt{2} < \alpha < \sqrt{2}$.

\section{Chiral Edge states}
In transport measurements through a mesoscopic two-dimensional system, 
the pair of edges parallel to the longitudinal current offers a sharp confining 
potential to the charge carriers in the transverse direction. As a result, 
the bulk 2D bands decompose into several 1D bands (or sub-bands) whose propagation 
vector is restricted along the longitudinal direction. The transport coefficients 
of a finite system is hence controlled by these sub-bands. The wave functions associated 
with these bands may be spread out in the bulk of the sample or localized at the edges.

\begin{widetext}
\begin{figure*}[htbp]
\includegraphics[trim={1cm 0cm 0cm  0cm},clip,width=18cm]{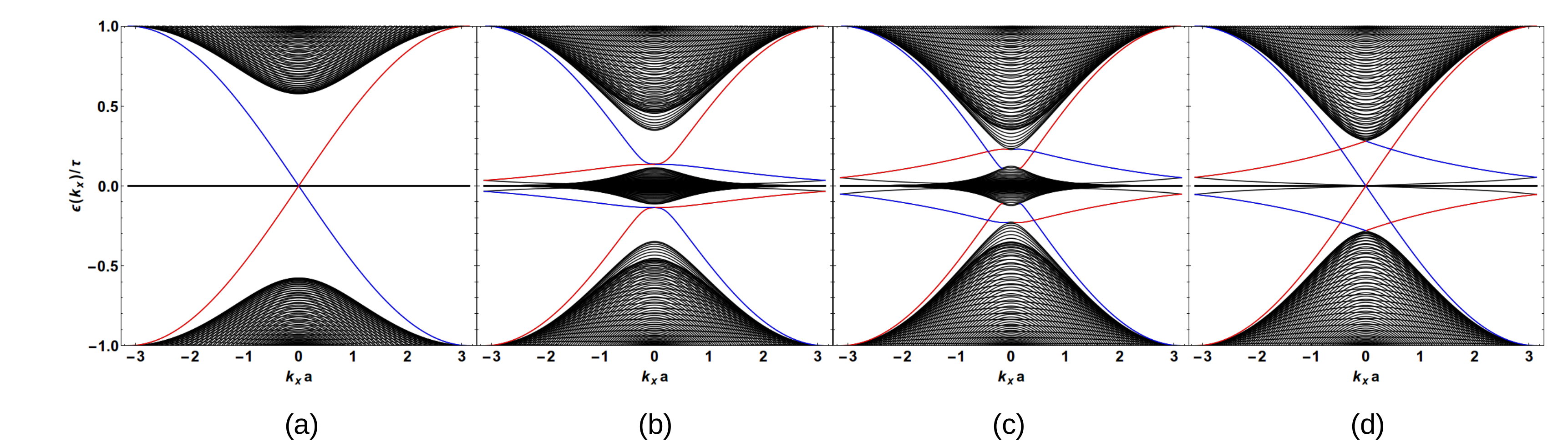}
\caption{Radiation-dressed band structure of  $\alpha$-$\mathcal{T}_3$ nanoribbon
with armchair edges for (a) $\alpha=0$, (b) $\alpha=0.5$, (c) $\alpha=0.8$ and
(d) $ \alpha=1.0$. The red/blue curves represent edge states propagating
on top/bottom edges, while the black curves represent the bulk bands. }
\label{edge-framed}
\end{figure*}
\end{widetext}

The bulk-boundary correspondence (BBC)\cite{hatsugai} tells that chiral edge states 
appear at the boundaries of a band insulator, if the Chern number of the 2D bulk 
band is non-zero. The number of the chiral modes along an edge is equal to the 
Chern number of the bulk band. These edge states show up in the 1D band structure 
as channels bridging the gap between the bulk bands. Since the irradiated 
$\alpha$-$\mathcal{T}_3$ lattice is a Floquet-Chern insulator, it is expected to host 
topological edge states. We show the existence of chiral edge states by 
numerically computing the low-energy band structure of $\alpha$-$\mathcal{T}_3$ 
armchair nanoribbon  driven by the off-resonant radiation (Fig.\ref{edge-framed}). 
The following parameters have been used: width of the ribbon 
$\approx$ 130 $a$, $\tau/\hbar\omega= 0.1$ and $J_1(\eta) = 0.8$. 
The edge states denoted by blue and red curves are localized at opposite edges of 
the ribbon. The slope of the edge bands determine the group
velocity of the electronic states. Thus, the edge states move with opposite 
group velocities at the two edges. They form a connection between the gapped 
bulk bands (black ensemble of curves). There is only one gapless edge band for 
$ \alpha < 1/\sqrt{2} $ as shown in Fig. \ref{edge-framed}(a) and (b), 
consistent with the obtained Chern number shown in Fig. \ref{singularity}. 
On the other hand, Fig. \ref{edge-framed}(c) and (d) display
that there are two edge bands for $\alpha =0.8$ and $\alpha= 1$, which
agrees with the calculated Chern number $\mathcal{C}$ =2. The BBC principle 
holds true in all these observations.

\section{Summary and Conclusion}
We have considered $\alpha$-$\mathcal{T}_3$ lattice illuminated by intense circularly
polarized radiation of frequency much higher than the bandwidth of the system. 
Using the off-resonant approximation, we have derived exact analytical 
expressions of the effective static band structure over the full Brillouin zone.
It is observed that the triple point degeneracy is completely lifted 
due to the broken TRS symmetry caused by circularly polarized light.
It leads to unequal photoinduced gaps at ${\bf K}$ and ${\bf K}^{\prime}$ (except 
for monolayer graphene and dice lattice) due to the lack of inversion symmetry.
At $\alpha= 1/\sqrt{2}$, the semimetalic phase is restored 
due to closing of gap between flat and valence bands at ${\bf K}$ and that between 
the conduction and flat bands at ${\bf K}^{\prime}$.
The low-energy Dirac cones at ${\bf K}$ and  ${\bf K}^{\prime}$ points resurface
at the gap closing point. The gap-closing value of $\alpha$ is insensitive to the radiation amplitude and polarization of light (except linear) within the off-resonant approximation. 
The $\alpha$-$\mathcal{T}_3$ lattice illuminated by the circularly polarized radiation 
is transformed to a Haldane-like Chern insulator.
We find that there is a topological phase transition from the Chern number
$\mathcal{C}=1 (0,-1)$ to a Chern number $\mathcal{C}=2 (0,-2)$ at the band closing point,
where $\mathcal{C}$ is the Chern number of the valence (flat, conduction) band.
This is an example of a three-band system having larger Chern number. The 
effect of non-trivial topology of the system should get reflected in the 
transport measurements through the chiral edge channels as shown for 
the armchair configuration.

\begin{center}
{\bf ACKNOWLEDGEMENT}
\end{center}
We would like to thank Firoz Islam and Sonu Verma for useful discussion.

\end{document}